\documentstyle[12pt]{article} 

\begin{document} \begin{titlepage} 

\title{ \Huge The Triumph and Limitations of
\\ Quantum Field Theory \\ }

\author{ David J. Gross\footnote
{Present address: Institute For Theoretical Physics,
University of California; Santa Barbara, California}
\thanks{ Supported in part by the National Science
Foundation under Grant PHY90- 21984.}
\\     Princeton  University\\ Princeton,
New Jersey\\ }  

\maketitle \abstract{Talk presented at the conference ``Historical
and Philosophical Reflections on the Foundations of Quantum
Field Theory,'' at Boston University, March 1996.  It will
be
published in the proceedings of this conference.}

\vskip 2truein  \end{titlepage}

\section{The Triumph of Quantum Field Theory}    

Although the title of this
session is \lq\lq The Foundations of Quantum Field Theory",  I shall talk, not
of the foundations of quantum field theory (QFT), but of its triumphs and
limitations.	I am not sure it is necessary to  formulate  
the foundations of QFT, or even to define precisely what QFT  is. QFT is what
quantum field theorists do.  For  a practicing high energy physicist,  nature is
a surer guide as to what quantum field theory is as well to what might supersede
it, is than the consistency of its axioms.  

	Quantum Field Theory (QFT) is today at a pinnacle of success. It provides  the
framework for the the standard model,  a theory  of all the observed  forces of
nature. This theory  describes the forces of electromagnetism, the weak
interaction  responsible for radioactivity,  and the strong nuclear force that
governs the structure of nuclei, as consequences of   local (gauge) symmetries.
These forces act on the fundamental constituents of matter, which have been
identified as pointlike quarks and leptons. The theory agrees astonishingly well
with experiment to an accuracy of $10^{-6}-10^{-10}$ for electrodynamics, of
$10^{-1}-10^{-4}$ for the weak interactions and of $1-10^{-2}$ for the strong
interactions. It has been tested down to distances of $10^{-18}$ cm. in some
cases. We can see no reason why QFT should not be adequate  down to distances of
order the  Planck length of  $10^{-33} $cm. where gravity becomes important. If
we note that classical electrodynamics and gravity  are simply limiting  cases
of their quantum field theoretic generalizations,  then quantum field theory
works from the Planck length to the edge of the universe--over $60$ orders of
magnitude. No other theory has been so universally successful. 

\subsection{The
Problems of the Past} 

It is hard for today's generation to remember the
situation  $35$ years ago, when field  theory had been abandoned by almost all
particle physicists. The exhilaration following the      development of Quantum
Electrodynamics (QED)   was short lived when the  same methods were applied to
the  new field of mesons and nucleons. The proliferation of   elementary
particles and their	 strong couplings,   as well as the misunderstanding 
and   discomfort with renormalization  theory, gave rise to despair with field
theoretic models  and to the conclusion  that quantum field theory  itself was
at  fault.

Renormalization was originally a response to the ultraviolet divergences that
appeared in the calculations of radiative corrections  in QED in a perturbative
expansion in the fine structure constant.  The basic observation was that  if
physical observables were 
expressed, not in terms of the bare parameters that  entered the definition of
the theory (like the bare mass of the electron)   and  refer  to idealized
measurements at infinitely small distances,  but rather in terms of the physical
parameters that are actually measurable at finite distances, then they would be
finite, calculable functions of these. Feynman, Schwinger, Tomanaga and Dyson
set forth the procedure for carrying out this  renormalization to all orders in
perturbation theory and proved that it yielded well-defined, finite results.
Even though this   program was very successful many physicists were
uncomfortable with renormalization, feeling that it was  merely a trick that
swept the fundamental problem of ultraviolet divergences under the  rug.

Furthermore, there was great concern as to the consistency of quantum field
theory at short distances.  Most four dimensional quantum field theories are not
asymptotically free, thus their short distance behavior is governed by strong
coupling and  thus not easily treatable. In the fifties it was suspected,
especially by Landau and his school, that the nonperturbative ultraviolet
behavior of QFT meant that these theories were inherently inconsistent,  since
they  violated unitarity (which means   that the total sum of probabilities  for
the  outcomes of measurements of some physical processes was not unity). This is
probably the case  for most non-asymptotically free theories, which are most
likely inconsistent as complete quantum field theories. The discovery of
asymptotic freedom, however, has provided  us with theories whose ultraviolet
behavior is    totally under  control. 

The disillusionment with QFT as a basis for the  theory of elementary particles
was also premature. What one was missing were many ingredients, including the
identification of the underlying gauge symmetry of the weak   interactions,  the
concept of spontaneous symmetry breaking that could explain how this symmetry
was hidden,  the  identification of the fundamental constituents of  the
nucleons as colored quarks, the  discovery of asymptotic freedom    which
explained how the elementary colored constituents  of hadrons could be seen at
short distances yet evade  detection through confinement, and the identification
of the underlying gauge symmetry of the strong interactions. Once these were
discovered,    it was but a short step to the construction of the standard
model, a gauge theory modeled on QED, which opened the  door  to the
understanding of mesons and nucleons.

\section{ The Lessons of Quantum Field Theory}   

The development  and successes
of QFT have taught us much about nature and the language we should use to
describe it. Some of the  lessons we have learned may transcend QFT. Indeed they
might point the way beyond QFT.  The most important lessons, in my opinion,
have to do with  symmetry principles and with the renormalization group.

\subsection{Symmetry}

The most important  lesson that we have learned  in this century it is that the
secret of nature is symmetry.      Starting with relativity, proceeding through
the development of quantum mechanics and culminating,  in the standard model
symmetry principles have assumed a central position  in the fundamental theories
of nature. Local gauge symmetries 
provide the basis of  the   standard model and of Einstein's   theory of
gravitation. 

Global symmetry principles   express the invariance of physical laws under  an
actual  transformation of the the   physical world.  Local  symmetry principles
express the invariance of   physical phenomena under a transformation of  our
description of them,  yet local symmetry  underlies dynamics. As Yang has
stated: {\sl Symmetry dictates interaction.} The first example  of a gauge
theory   was general relativity where diffeomorphism  invariance of spacetime
dictated  the laws of gravity.  In the standard model,  non-Abelian gauge
symmetry dictates the electroweak and strong forces. Today we believe that
global symmetries are unnatural. They smell of action at a distance.  We now
suspect, that   all fundamental symmetries are local gauge symmetries.  Global
symmetries  are either   broken, or  approximate, or they are the  remnants of
spontaneously broken local symmetries. Thus,  Poincar\`{e} invariance can be
regarded as the residual symmetry of  general relativity in the Minkowski vacuum
under  changes of the spacetime coordinates.

The  story of symmetry does not  end with gauge symmetries. In recent years we
have discovered a new and extremely powerful new symmetry---supersymmetry---
which might explain many mysteries of the standard model. We avidly await the
experimental discovery of this symmetry.  The search for even newer symmetries
is at the heart of many current attempts to go beyond the standard model. String
theory, for example,  shows signs of containing totally new and mysterious
symmetries with   greater  predictive  power. 

Another part  of the lesson of symmetry is that,  although the  secret of nature
is symmetry, much of the  texture of the world is due to mechanisms of symmetry
breaking. The spontaneous symmetry breaking of global and local gauge symmetries
is a  recurrent theme in  modern theoretical physics. In  quantum mechanical
systems with a finite number of degrees of freedom  global symmetries are
realized in only one way. The laws of physics are invariant and the ground state
of the theory is unique and symmetric. However, in systems with an infinite
number of degrees of freedom a second realization of symmetry    is possible, in
which    the ground state is asymmetric. This  spontaneous symmetry breaking is
responsible for magnetism, superconductivity, the structure of the unified
electro-weak theory and more. In such a situation the symmetry of nature's laws
is hidden from us.  Indeed,  the search for new symmetries of nature is based on
the possibility of finding mechanisms, such as spontaneous symmetry breaking or
confinement, that hide the new symmetry.   

There are two corollaries of the lesson of symmetry that are  relevant to our
understanding of QFT. First is   the importance of special quantum field
theories. A  common strategy  adopted years ago, say in constructive field
theory, was to consider theories  with only scalar  fields. Their study, it was
thought,  would  teach   us  the general principles of QFT and illuminate its
foundations. This,   to an extent, was achieved. But, in the absence of vector
or fermionic fields   one cannot  construct  either  gauge invariant  or
supersymmetric theories, with all of their special and rich phenomena. Today it
might be equally foolhardy to ignore quantum gravity in the  further development
of QFT. Indeed the fact that QFT  finds it so difficult to incorporate the
dynamics of spacetime suggests that we might search for more special theories.

Second, we have probably  exhausted  all possible symmetries of QFT. To find new
ones we need a richer framework.  Traditional quantum field theory is based on
the principles of locality and causality, on the principles of quantum mechanics
and on the principles of symmetry. It used to be thought that QFT, or even
particular quantum field theories, were the  unique way of realizing such
principles. String theory  provides us with an example of   a theory that
extends quantum field theory, yet embodies these same principles. It appears to
contain new and strange symmetries that do not appear in QFT.   If there are
new  organizing  principles of nature,  the framework of QFT may    simply not
be   rich enough. We may need string theory, or even more radical theories, to
deal with new symmetries, especially those of spacetime. 

\subsection{ The   Renormalization Group }               

The second important
lesson we have learned is  the idea of the renormalization group and effective
dynamics.   The decoupling of physical phenomena at different scales of energy
is an   essential characteristic of nature.   It is this feature of nature that
makes it possible to understand a	limited range of physical phenomena without
having to 
understand everything at once. The renormalization group describes the change of
our description of physics as we 
change the scale at which we probe nature. These methods are  especially
powerful in QFT which asserts control over physics at all scales.
Quantum field theories are most naturally formulated at short distances, where
locality can be most easily imposed, in terms of some fundamental  dynamical
degrees of freedom (described by quantum fields).  Measurement, however, always
refers to physics at some finite distance. We  can  describe the low energy
physics  we are interested in by  deriving an effective theory which involves
only the low momentum modes of the theory.  This  procedure, that   of {\it
integrating  out the high momentum  modes of   the quantum fields},   is the
essence of the renormalization group, a transformation   that  describes the
flow of couplings in   the   space of quantum field  theories as we reduce the
scale of energy.

	The characteristic behavior of the solutions of the renormalization group
equations is that they approach a finite dimensional sub-manifold in the
infinite dimensional space of all theories. This defines  an  {\it effective}
low energy theory, which is formulated  in terms of a finite number of degrees
of freedom and   parameters and is largely independent of the high energy
starting point. This effective low energy theory might be formulated in terms of
totally different quantum fields, but  it is equally {\it fundamental} to the
original high energy formulation, insofar as our only concern is low energy
physics.  

Thus, for example,  QCD is the theory of quarks  whose   interactions are
mediated by gluons. This is the appropriate description at energies of billions
of electron volts. However, if we wish to describe the properties of ordinary
nuclei, at energies of millions of electron volts  we employ instead an {\it
effective theory} of nucleons, composites of the quarks, whose   interactions
are mediated by other quark composites---mesons. Similarly,  in order to discuss
the properties of  ordinary matter made of atoms at energies of a	few electron
volts we  can treat the nuclei as pointlike 
particles, ignore their internal structure and take into account only the
electromagnetic interactions of the charged nuclei and electrons.  
The renormalization group influences the way we think  about  QFT itself. One
implication is that there may be  more than one, equally fundamental,
formulation  of a particular QFT; each  appropriate for describing physics at a
different scale of energy. Thus, the formulation of QCD as a theory of quarks
and gluons is  appropriate at high energies where, due to asymptotic freedom,
these degrees of freedom are weakly coupled. At low  energies it is quite
possible, although not yet realized in practice, that the theory is equivalent
to a theory of  strings--- describing mesons as   tubes of confined
chromodynamic flux. Both formulations might be equivalent and complete, each
appropriate to a different energy regime. Indeed, as this example suggests, a
quantum field theory might be equivalent to a totally different kind of theory,
such as a string theory.     The renormalization group has had a profound
influence on how we think about renormalizability. Renormalizability was often
regarded a selection principle for  QFT.   Many quantum field theories, those
whose couplings  had   dimensions of powers of an inverse mass (such as  the
Fermi theory of weak interactions),  were not renormalizable.  This meant that,
once such interactions were introduced,  it was necessary  to  specify  an
infinite number of additional   interactions with  an infinite number of  free
parameters in order    to ensure  the finiteness of physical observables.   This
seemed  physically nonsensical, since  such a theory has no predictive power and
was taken to be the reason why theories of nature, such as QED, were described
by renormalizable quantum field theories.

Our present view of things is quite different. The renormalization group
philosophy   can be applied to the standard model itself.  Imagine that we have
a unified theory   whose characteristic energy scale, $\Lambda$,  is very large
or whose characteristic distance scale, $\hbar c  /\Lambda$,  is very small (say
the Planck length of $10^{33}$cm.).  Assume further that just below this scale
the  theory can be expressed in terms of local field variables.  As to what
happens at the unification scale itself we  assume  nothing, except that just
below this scale the theory can be described  by  a local quantum field theory.
(String theory does provide us with an example of such a unified theory, which
includes gravity  and can be expressed by local field theory at distances much
larger than the Planck length.) Even in the absence of knowledge regarding the
unified theory,  we can determine  the most general quantum field theory. In
absence of knowledge as to the principles of unification  this theory  has an
infinite number of  arbitrary parameters describing  all   possible fields and
all possible interactions.   We also   assume that all the  dimensionless
couplings that characterize the theory at energy $\Lambda$ are of   order one
(what else could they be?).   Such a theory is useless to describe the physics
at high energy, however, at low energies, of order $E$, the effective dynamics,
the effective Lagrangian  that describes physics up to corrections of order
${E/\Lambda }$,  will be parameterized by a finite number of couplings. The
renormalization group describes  how the various couplings run with energy.  We
start at $\Lambda$ with whatever the final unified theory and then one can show
that  the low energy physics will be described by the most general
renormalizable field theory consistent with the assumed 
couplings plus  non-renormalizable interactions that are suppressed by powers of
the energy relative to the 
cutoff.  If we demand further that the theory at the 
scale $\Lambda$ contain the local gauge symmetry that we observe in nature, then
the effective low energy theory will be described by    the standard model  up
to terms that are negligible by inverse powers of the large scale compared to
the energy that we observe.  The extra interactions   will give rise to weak
effects, such as  gravity  or baryon decay. But these are  very small and
unobservable at low energy.

 Non-renormalizable theories were once rejected since, if they had  couplings of
order one at low energies, then their high energy behavior was uncontrollable
unless one specified  an infinite number of arbitrary parameters. This is now
turned around. If all couplings are moderate at high energies, then
non-renormalizable interactions are unobservable at low energies.   Furthermore,
the standard model is   the inevitable consequence of any unified theory, any
form of the final theory, as long as it is local at the very high energy scale
and contains the observed low energy symmetries.   In some sense   this is
pleasing, we understand why the standard model emerges at low energy.  But  from
the point of view of the unified theory that surely awaits us at very high
energy it is disappointing, since our low energy theory tells us little about
what the final theory can be. Indeed, the high energy theory need not be a QFT
at all.

\eject 
\section{QCD As A Perfect  QFT}         

For those who ever felt
uncomfortable with ultraviolet divergences, renormalization theory or the
arbitrary parameters  of quantum field theory,  QCD offers the example of a
perfect quantum field theory. By this I mean: 

\begin{itemize} 

\item This theory
has no ultraviolet divergences at all. The local ({\it bare}) coupling vanishes,
and the         only infinities that appear are due to the fact that one
sometimes expresses observables measured        at finite distances in terms of
those measured at infinitely small distances.  

\item	The theory has no free, adjustable parameters 
(neglecting the irrelevant   quark masses), and dimensional observables are
calculable in terms of the dynamically produced mass scale of the      theory
$m = \Lambda  \exp[ -1/g_0^2 ]$, where $g_0$ is the {\it bare} coupling that
characterizes the theory at  high energies of order $\Lambda$.

\item The theory shows no diseases when extrapolated to infinitely high
energies. To the contrary, asymptotic freedom means that at high energies   QCD
becomes simple and perturbation theory is  a better and better approximation.
\end{itemize}           

\noindent Thus, QCD provides the first example of a complete
theory with no adjustable parameters and with no indication within the theory of
a distance scale at which it must break down.

	There is a price to be paid for these wonderful features. The absence of
adjustable parameters means that there are no small parameters in the theory.
The generation 
of a dynamical mass scale means that perturbative methods cannot suffice for
most questions. The flip side of asymptotic      freedom is infra-red slavery,
so that the large distance properties of the theory, including the 
phenomenon of confinement, the dynamics of chiral 
symmetry breaking and the structure of  hadrons are issues of strong coupling.
What are the limitations of such a QFT? In  traditional  terms there are none.
Yet, even if we knew not of the electroweak and gravitational interactions, we
might suspect that the theory is incomplete. Not in the sense that it is
inconsistent,  but rather that there are questions that can be asked which it is
powerless to answer; such as  why is the gauge group $SU(3)$ or what dictates
the dynamics of spacetime?  

\section{  The Limitations of  QFT}

Quantum field theory is  a mature subject; the frontier of fundamental physics
lies elsewhere. Nonetheless there are many  open problems in quantum field
theory that  that should and will be  addressed   in the next decades.

First there are problems having to do with 
QCD, our most complete field theory. Much is understood, but much remains to be
understood. These problems include the proof of the existence of QCD and of
confinement; the development of analytic methods to control  QCD in the
infrared; and the development of numerical algorithms for Minkowski space and
scattering amplitudes.  The second class of problems are more general than QCD,
but would help in solving it as well. These include the development of large N
methods; the formulation of a non-perturbative continuum regularization;   the
rigorous formulation of renormalization flow in the space of Hamiltonians; a
first quantized path integral representation of gauge mesons and the  graviton;
the exploration of the phase structure of particular theories, particularly
supersymmetric gauge theories; the complete classification and understanding of
two-dimensional conformal field theories and integrable models; and  the
discovery and solution of special integrable  quantum field theories.  

		You will have noticed that the big problems of high energy physics are not on
the above  list. These include:  the unification of forces, the mass hierarchy
problem (namely why is the scale of the electroweak symmetry breaking and the
mass scale of the strong interactions smaller than the Planck or unification
scale by 14 to 18 orders of magnitude), the origin of lepton-quark families, the
explanation of the parameters of the standard model,  quantum gravity, the
smallness or vanishing of the cosmological constant, the early history of the
universe   \dots .		

The reason I have not listed these is that  
I	believe  that their resolution does not originate in 
quantum field theory at all. To solve these we will have to go beyond quantum
field theory to the next stage, for example to string theory. In this sense QFT
has reached true maturity. Not only   do we  marvel at its success but we are
aware of its boundaries. To truly understand a physical theory it is necessary
to have the perspective of the   next stage of physics that supersedes it. Thus
we understand  
classical mechanics much better in the light of quantum mechanics,
electrodynamics much better after QED.  Perhaps the true understanding of QFT
will only transpire  after we find its successor.

The search for a replacement for QFT has been going on ever since its invention.
Every conceptual and technical difficulty that was encountered was taken as
evidence for a fundamental length at which QFT breaks down. With the  success of
QFT as embodied in the standard model the search for a fundamental length has
been pushed down  to the Planck length. There  almost everyone believes that a
new framework will be required, since many of the basic concepts of QFT are
unclear once space-time fluctuates violently. The  longstanding problem of
quantizing gravity is probably impossible within the framework of quantum field
theory. Einstein's theory of gravity  appears to be an effective 
theory, whose dimensional coupling, Newton's constant $G_N$,  arises from the
scale of unification  which might be  close to the Planck mass $M_p$, i.e.,
$G_N\propto 1/M_p^2  $. General relativity is then  simply an   incredibly weak
force that  survives  at low energies and is only observable  since it couples
coherently,  via  long range forces, to  mass, so that we can observe its
effects on  large objects.  QFT has proved useless in incorporating quantum
gravity into a consistent theory at the Planck scale.  We need to go beyond QFT,
to a theory of strings or to something else, to describe quantum gravity.

There are other indications of the limitations of QFT. The  
very success of QFT in providing us  with an extremely successful theory of all
the non-gravitational forces of nature has made  it clear that this framework
cannot explain many of the    features and  parameters of the standard  model
which cry out for explanation. In the days before the standard model it was
possible to believe that the requirement of renormalizability or symmetry would
be sufficient to yield  total  predictive power (as in QCD). But today's
understanding   makes it clear that these principles are not sufficient.
Thus,  the limitations of QFT are   not those of consistency or incompleteness
in its own terms,  but rather of insufficiency and incompleteness in broader
terms. If we restrict ourselves to effective field theories, or the use of QFT
in dealing with nonrelativistic many body systems, or fundamental theories of
limited domain (such as QCD), then QFT is in fine shape.  But if  we are to come
to grips with the  quantization of  the dynamics of spacetime then QFT is, I
believe,  inadequate. I also believe that we will learn  much about QFT itself
from its successor. For example, there are certain features of special quantum
field theories (such as the recently developed duality symmetries)  whose deeper
understanding might require the embedding of field theory within string theory.

\section{  Beyond  QFT---String Theory} 

We have one strong candidate for  an extension of physics beyond  QFT that does
claim to be able to answer the  questions  that QFT cannot and more---string
theory. String theory is a radically conservative extension of the principles of
physics, in which one  introduces fundamental degrees of freedom that are not
pointlike, but rather have the structure of extended one-dimensional
objects---strings, 
while leaving untouched (at least in the beginning) the other principles of
causality, relativistic invariance and quantum mechanics. The structure of this
theory, which appears to be rather unique and free of  any non-dynamical
parameters,  is quite remarkable. It yields a consistent theory of  quantum
gravity, at least in the perturbative, weak field domain, providing us with an
existence proof that gravity and quantum mechanics are mutually consistent. In
addition, it  appears to  possess all the ingredients that  would be necessary
to reproduce and explain the standard model.  Most important, it is definitely
an extension of the conceptual framework of physics beyond QFT.     

	There have been two major revolutions completed in this century: relativity,
special and general, and quantum mechanics. These were associated with two of
the three dimensional parameters of physics: $\hbar$, Planck's quantum of
action, and  ${  c}$, the velocity of light. Both involved major conceptual
changes in the framework of physics, but reduced to  classical non-relativistic
physics when $\hbar$ or $1\! /\!c $ could be regarded as small.   The last
dimensional parameter  we need in order to establish a set of fundamental
dimensional units of nature,  is Newton's gravitational constant,  which  sets
the 
fundamental (Planck) scale of length or energy.  Many  of us  believe that
string theory is the revolution  associated with this  last of the dimensional
parameters of nature.  At large distances,  compared to the string  scale of
approximately  $10^{-33}$cm.,  string theory goes over into field theory.  At
shorter distances it is bound to be very different, indeed  it calls  into
question what we mean by distance or spacetime itself.  

 The reason we are unable to construct predictive models based on string theory
is our lack of understanding of the nonperturbative dynamics  of string theory.
Our present understanding of string theory is very primitive.  It appears to be
a  totally consistent theory, that does away with pointlike structures and
hints at a fundamental revision of the notions of space and time at short
distances while at  the same time reducing to field theory at large distances.
It introduces  a fundamental length in a way that had not  been envisaged---not
by, for example, discretizing space and time---but rather by replacing the
fundamental point-like constituents of matter with  extended, non-local
strings. The constituents are non-local but they interact locally;   this is
sufficient to preserve the  usual consequences of locality---causality as
expressed in the analyticity of scattering amplitudes. 

To be more specific, string theory is constructed to date by the method of {\it
first quantization}.  Feynman's approach to QFT, wherein scattering amplitudes
are constructed by summing over the trajectories of particles, with each
history weighted by the exponential of ($i$ times) the classical action   given
by the proper length of the of the path, is generalized to strings by replacing
the length of the particle trajectory with the area swept out by the string as
it moves in spacetime. This yields a perturbative expansion of  the amplitudes
in powers of the string coupling, which is analogous to the Feynman diagram
expansion of QFT. However, string theory exhibits    profound differences  from
QFT. First,  there is no longer any ambiguity, or freedom, in specifying the
string interactions, since there is no longer an invariant way of specifying
when and where the interaction took place.  Consequently, the string coupling
itself becomes a dynamical 
variable, whose value should ultimately  be determined (in ways we do not yet
understand). Furthermore, there are only a few, perhaps  only one, consistent
string theory. Finally, the issue of ultraviolet divergences is automatically
solved, the smoothing out  of world-lines to world-tubes renders the
interactions extremely soft and ensures that string amplitudes are totally
finite.

At low energies string theory goes over into field theory. That means that we
can describe the scattering amplitudes by an effective field theory describing
the light particles to any degree of approximation in powers of the momenta
$p/M_{ \rm planck}$. However  string theory is not just a complicated field
theory. It exhibits features at short distances or high energies that are
profoundly different than QFT; for example the Gaussian falloff of scattering
amplitudes at large momenta. At the moment we are still groping towards an
understanding of its properties for strong coupling and for short distances.
In our  eventual  understanding of string theory we might  have to undergo a
discontinuous conceptual change in the way we look at the world similar to that
which occurred in the development of relativity and quantum mechanics.  I think
that we are in some sense in a situation analogous to where physics was  in the
the beginning of the development of quantum mechanics, where one had a
semiclassical approximation to quantum mechanics, that was not yet part of a
consistent, coherent framework.  There was an enormous amount of confusion until
quantum mechanics was finally discovered.

What will this revolution lead to?
Which of our concepts will have to be modified? There are many hints that  our
concepts of spacetime, which are so fundamental to our understanding of nature,
will have to be altered.  The first hint is based on a stringy analysis of the
measurement of position, following Heisenberg's famous analysis in the quantum
mechanics. Already in ordinary quantum mechanics  space becomes somewhat fuzzy.
The very act of measurement of the position of a particle can change its
position.  In order to perform a measurement of position $x$, with  a small
uncertainty of order $\Delta x$, we require probes of very high energy $E$. That
is why we employ microscopes with high frequency (energy) rays or particle
accelerators to explore short distances. The precise relation is that $$\Delta
x\approx {\hbar c \over E}, $$ where $\hbar $ is Planck's quantum of action and
$c$ is the velocity of light. In string theory, however, the probes themselves
are not pointlike, but rather extended objects, and thus there is another
limitation as to how precisely we can measure short distances. As energy is
pumped into the string it expands and thus there is an additional uncertainty
proportional to the energy. All together $$\Delta x\approx {\hbar c \over E} +
{G E \over c^5}. $$ Consequently  it appears impossible to measure distances
shorter than the Planck length. 

The second hint is based on  a symmetry of string theory known as duality.
Imagine a string that  lives in a world in which one of the spatial dimensions
is a little circle of radius $R$. Such  situations are common in string theory
and indeed necessary if we are to reconcile the fact that the string theories
are naturally formulated in nine spatial dimensions so, that if they are to look
like the real world, six dimensions must be curled up, {\sl compactified}, into
a small space. Such perturbative solutions of realistic string theories have
been found and are the basis for 
phenomenological  string models. Returning to the simple example of a circle,
duality states that the theory is identical in all of its physical properties to
one that is compactified on a circle of radius $\bar R=L_p^2/R$, where $L_p$ is
the ubiquitous Planck length of $10^{-33}$cm. Thus if we try to make the extent
of one of the dimensions of space very small, by curling up one dimension into a
circle of very small radius  $R$, we would instead interpret this as a  world in
which the circle had  a very large radius $\bar R$. The minimal size of the
circle is of order $L_p$. This  property is inherently stringy. It arises from
the existence of stringy states that wind around  the spatial circle and again
suggests that spatial dimensions less  the Planck length have no meaning.

Another threat to our conventional view of spacetime is the discovery  that in
string theory the very topology of space-time can continuously be altered. In
perturbative string theory  there are families of solutions labeled by various
parameters. In some cases these solutions can be pictured as describing strings
propagating on a certain curved spatial manifold. As one varies the parameters
the shape and geometry of the background manifold varies. It turns out that by
varying these parameters one can continuously deform the theory so that  the
essential geometry of the background manifold changes. Thus one can go smoothly
from a string  moving in one geometry to a string  moving in another; although
in between there is no simple spacetime description. This phenomenon  cannot be
explained by ordinary quantum field theories.

Finally,  during this last year, new developments  have been made in the
understanding of the structure  of string theory.  A remarkable set of
conjectures have been formulated and tested that relate quite different string
theories to each other (S,T,U dualities) for different values of their
parameters and  for  different background spacetime manifolds.    Until now  the
methods we have employed to construct string theories have been quite
conservative. To calculate string scattering amplitudes  one used the method of
\lq\lq first quantization", in which  the amplitudes are constructed by summing
over path histories of propagating strings, with  each path weighted by the
exponential of the classical action (the area of the world sheet swept out by
the string as it moves in spacetime. This approach, originally developed by
Feynman for QED, is quite adequate for perturbative calculations. It was
envisaged that to do better one would, as in QFT, develop a string field theory.
However, these new developments suggests that in addition to stringlike objects,
\lq\lq string theory" contains other extended objects of higher internal
dimension, that cannot be treated by  the same first quantized methods and for
which this approach is inadequate. Even stranger, some of the duality symmetries
of string theory  connect theories  whose   couplings ($g_1$ and $g_2$)    are
inversely related  $g_1 = 1/g_2$. This is  a generalization of electrodynamic
duality,  wherein the electric and magnetic fields  and theire charges ($e$ and
$g$, related by the Dirac quantization condition $eg = 2 \pi \hbar $) are
interchanged. 

These developments  hint  that the ultimate formulation of string theory will be
quite different that originally envisaged. It might be one in which strings   do
not play a fundamental role and  it might be a theory that cannot be
constructed  as the quantization of a classical theory. Thus it appears that we
are headed for a real theoretical crisis in the development of string  theory. A
welcome crisis, in my opinion, one that could force us to radically new ideas.

\subsection{Lessons for QFT}

What can we learn from  string theory  about quantum field theory? There are a
few lessons         that we can already extract, and I expect  that many more
will emerge in the future. 

\begin{itemize} 

\item{ }     First this theory, used
simply as an example of a unified theory at a very high energy scale, provides
us with a vindication of the modern  philosophy of the  renormalization group
and the  
effective Lagrangian that I discussed previously.  
Using string theory as  the  theory at the cutoff we can verify that at energies
low compared to the cutoff (the Planck mass) all observables can be reproduced
by an effective local quantum field theory and, most importantly, all
dimensionless couplings in this effective, high-energy theory, are of the  same
order of magnitude.

		Thus, we have an  example of a theory, which as far as we can see, is
consistent at arbitrarily high energies and reduces at low energy to quantum
field theory. String theory could  explain the emergence of quantum field theory
in the low energy limit,  much as quantum       mechanics explains classical
mechanics, whose equations can be understood as determining  the saddlepoints of
the  quantum path integral in the limit of small $\hbar$.

\item{ } We also learn that  the  very same quantum field theories that play   a
special role in nature are those that emerge from the string. These include
general relativity,  non-Abelian gauge theories  and (perhaps) supersymmetric
quantum field theories. Thus, string theory could explain the distinguished role
of these theories. From a practical point of view string theory can be used to
motivate the construction of novel field theoretic models that include  less
familiar interactions  as well as new kinds of particles, such as  axions or
dilatons,  that are ubiquitous features of low energy string physics. 

\item{ } Finally, it appears that some of the new and mysterious dualities of
supersymmetric field theories have their natural explanation in the framework of
string theory. To truly understand these features of QFT it may be necessary to
consider a field theory as part of a string  theory and use the latter to
understand the former.
 \end{itemize}

\section{Conclusions:}

	I believe that we are living in revolutionary times, where many of the basic
principles of physics are being 
challenged by the need  to go  beyond QFT and in which many of our basic
concepts will  require fundamental revision.  Will the exploration of the
foundations or  the  philosophical meaning of QFT help  us in these tasks?   I
admit that my prejudice is that the answer is no. The issues that face us now
have little  to do with those that were confronted in the struggle to make sense
of QFT. Rather, it is the surprises that  we unearth in the experimental
exploration of nature as well as the  those that emerge in the theoretical
exploration of our emerging theories that will force us to radical modifications
of our basic preconceptions.

\end{document}